\documentclass[twocolumn,showpacs,preprintnumbers,prb,aps,superscriptaddress]{revtex4-2}
\usepackage{amsmath}
\usepackage{amssymb}
\usepackage[retain-unity-mantissa=false]{siunitx}
\usepackage{graphicx}
\usepackage{xcolor}
\usepackage[caption=false]{subfig}
\usepackage{placeins}
\usepackage[colorlinks=true, linkcolor=blue, citecolor=blue, filecolor=blue, urlcolor=blue,pdfproducer={.},pdfcreator={.}]{hyperref}
\usepackage{cleveref}
\usepackage{orcidlink}
\usepackage{academicons}    % für das ORCID-Symbol
\usepackage{blindtext}
\usepackage{comment}
\usepackage{tikz}%tikz ist kein Zeichenprogramm (tikz is not a drawing program). Program to draw pictures
\usepackage{tikz-3dplot} %makes it 3D!
\usetikzlibrary{calc}		%tikz now can calulcate positions itself
\usetikzlibrary{positioning}			%
\usetikzlibrary{backgrounds}			%
\usetikzlibrary{decorations.pathmorphing}	% looks nicer
\usepackage{verbatim}
\usepackage{ulem}
\usepackage{physics}
\usepackage[commandnameprefix=ifneeded, commentmarkup=uwave]{changes}

\makeatletter
\def\@fnsymbol#1{\ensuremath{\ifcase#1\or *\or \ddagger\or
   \mathsection\or \mathparagraph\or \|\or **\or \dagger\dagger
   \or \ddagger\ddagger \else\@ctrerr\fi}}
\makeatother

\definechangesauthor[name=Chris, color=orange]{C}
\definechangesauthor[name=Baerbel, color=magenta]{B}

\newcommand{\niceint}[2]{\int \mathrm{d}#1 #2 } %nicer looking integral
\renewcommand{\eqref}[1]{\cref{#1}}
\newcommand{\e}[1]{{\mathrm{e}}^{#1}} % exp function with real exponent

\newcommand{\nm}{\nano\meter}

\newcommand{\orcid}[1]{%
    \href{https://orcid.org/#1}{\textcolor{black}{\aiOrcid}}%
}

\begin{document}
	\title{Thermalization of optically excited Fermi systems:\\ electron-electron collisions in solid metals}
	\author{Stephanie \surname{Roden} \orcidlink{0009-0000-4426-3644}}
    \email{roden@rptu.de}
    \author{Christopher Seibel \orcidlink{0000-0003-1513-1364}}
    \author{Tobias Held \orcidlink{0009-0009-8925-1810}}
    \author{Markus Uehlein \orcidlink{0000-0002-3193-3749}}
    \author{Sebastian T. Weber \orcidlink{0000-0002-9090-2248}}
    \author{Baerbel Rethfeld \orcidlink{0009-0008-9921-4127}}    \email{rethfeld@rptu.de}
	\affiliation{Department of Physics and Research Center OPTIMAS, RPTU University Kaiserslautern-Landau, 67663 Kaiserslautern, Germany}
	
\begin{abstract}
Ultrafast optical excitation of metals induces a non-equilibrium energy distribution in the electronic system,
with a characteristic step-structure determined by Pauli blocking.
On a femtosecond timescale, electron-electron scattering drives the electrons towards a hot Fermi distribution.
In this work, we present a derivation of the full electron-electron Boltzmann collision integral within the random-$\boldsymbol{k}$ approximation.
Building on this approach, we trace the temporal evolution of the electron energy distribution towards equilibrium, for an excited but strongly degenerate Fermi system. 
Furthermore, we examine to which extent the resulting dynamics can be captured by the numerically simpler relaxation time approach, applying a constant and an energy-dependent relaxation time derived from Fermi-liquid theory.
We find a better agreement with the latter, while specific features caused by the balance of scattering and reoccupation can only be captured with a full collision integral.
\end{abstract}
	
	\date{\today}
	
	\maketitle

    \section{Introduction}
 Femtosecond laser pulses constitute a cornerstone of modern solid-state physics, as they enable the manipulation of matter on timescales intrinsic to its fundamental electronic and structural processes \cite{Strickland1985, Fann1992b, Krausz2009, Bauer2015}.
In optical pump-probe experiments, a material is first excited by an ultrashort laser pulse, which increases the energy of the electronic subsystem and drives it out of equilibrium.
Many key quantities characterizing the electron system cannot be probed directly and must instead be inferred, for example, from the optical response.
However, the optical response depends not only on the total energy or density of the system but also on how that energy is distributed among the electrons \cite{Lindhard1954,Oppeneer2004,Ndione2024prb, Seibel2025CommPhys}.
Moreover, the transient electron distribution directly affects material properties, such as the optical response and the transport of particles and energy.
A detailed understanding of the non-equilibrium electron energy distribution is therefore essential for the proper interpretation of optical measurements.

Following the excitation by a laser pulse, the electrons undergo two principal relaxation processes.
First, they redistribute energy among themselves and approach a higher-temperature Fermi distribution — a process referred to as thermalization.
Second, they transfer energy to the phonon system, leading to equilibration between the electronic and lattice subsystems.
For sufficiently strong excitation, thermalization occurs on a much shorter timescale than equilibration, allowing these processes to be considered separately.
Under such conditions, thermalization is predominantly mediated by electron-electron collisions.

In this work, we investigate the thermalization of a laser-excited, non-equilibrium electron system as it relaxes toward a hot Fermi distribution.
We assume optical excitation of moderate intensity, which preserves a strong degeneracy of the total system. 
We then characterize the resulting non-equilibrium state and observe how different descriptions of the electron-electron collisions lead to a different evolution of the energy distribution during the thermalization process. 
Finally, we determine the energy-dependent occupation time, which is significantly influenced by Pauli blocking effects. For a free-electron density of states, this time is almost the same for electrons above the Fermi energy level as it is for unoccupied states (holes) below it.
While a Fermi-liquid expression of single-electron scattering coincides well with the results of the full collision integral in distinct energy ranges, specific features are determined by the full ensemble dynamics and cannot be captured in a simplified approach.
    
    \section{Modelling the thermalization of electron systems\label{sec:Capturing_electron-electron_collisions}}
    In this section, we present different descriptions for the change of the electron energy distribution which can enter, e.g., the collision term of a Boltzmann equation. 
    We begin with a general expression and an analysis of the possible scattering terms in a spin-degenerate system. 
    Then, we derive the full electron-electron collision integral within the random-$\boldsymbol{k}$ approximation, considering spherically-averaged quantities and present the matrix element applied in this work.
    For comparison, we use the numerically much simpler relaxation time approach. We introduce an energy-dependent scattering frequency obtained from Fermi liquid theory, which captures effects arising in strongly degenerate Fermi system. 
    
    \subsection{Momentum-resolved collision integral}
    \label{subsec:el_el_collisions}
        \begin{figure}[]
            \centering
            \begin{tikzpicture}[scale=1.6]
                \coordinate (a_cl) at (-4,0);
            
                \draw [thick, black] ($(a_cl) + (-1,0.5)$) -- ($(a_cl)+(0,0.25)$) node [pos=0.3,above] {$\textbf{p}\, \sigma$};
                \draw [>=latex, ->, thick, color=black] ($(a_cl)+(0,0.25)$) -- ($(a_cl)+(1,0.5)$) node [pos=0.7, above] {$\textbf{p'}\, \sigma$};
                \draw [thick, black] ($(a_cl) - (1,0.5)$) -- ($(a_cl)-(0,0.25)$) node [pos=0.3,below] {$\textbf{k}\, \sigma$};
                \draw [>=latex, ->, thick, color=black] ($(a_cl)-(0,0.25)$) -- ($(a_cl) + (1,-0.5)$) node [pos=0.7, below] {$\textbf{k'}\, \sigma$};
                \draw [thick, densely dashed] ($(a_cl) +(0,0.25)$) -- ($(a_cl)-(0,0.25)$); 
            
                \coordinate (a_cr) at (-1.5,0);
                \draw [black] ($(a_cr)!0.5!(a_cl) - (0.075, 0)$) -- ($(a_cr)!0.5!(a_cl) + (0.075, 0)$);
            
                \draw [thick, black] ($(a_cr) + (-1,0.5)$) -- ($(a_cr)+(0,0.25)$) node [pos=0.3,above] {$\textbf{p}\, \sigma$};
                \draw [>=latex, ->, thick, color=black] ($(a_cr)+(0,0.25)$) -- ($(a_cr)+(1,-0.5)$) node [shift={(-0.3,0.1)}, below] {$\textbf{k'}\, \sigma$};
                \draw [thick, black] ($(a_cr) - (1,0.5)$) -- ($(a_cr)-(0,0.25)$) node [pos=0.3,below] {$\textbf{k}\, \sigma$};
                \draw [>=latex, ->, thick, color=black] ($(a_cr)-(0,0.25)$) -- ($(a_cr) + (1,0.5)$) node [shift={(-0.3,-0.1)}, above] {$\textbf{p'}\, \sigma$};
                \draw [thick, densely dashed] ($(a_cr) +(0,0.25)$) -- ($(a_cr)-(0,0.25)$); 
                \node at ($(a_cl) + (-1.1,1.0)$) {a)};
                \node at ($(a_cl)!0.5!(a_cr) + (0,1.0)$) {spin conserving, same spin};
            \end{tikzpicture}
            \begin{tikzpicture}[scale=1.5]
                \coordinate (b) at ($(a_cl) + (0,-0.25)$);
                
                \draw [thick, black] ($(b) + (-1,0.5)$) -- ($(b)+(0,0.25)$) node [pos=0.3,above] {$\textbf{p}\, \sigma$};
                \draw [>=latex, ->, thick, color=black] ($(b)+(0,0.25)$) -- ($(b)+(1,0.5)$) node [pos=0.7, above] {$\textbf{p'}\, \sigma$};
                \draw [thick, black] ($(b) - (1,0.5)$) -- ($(b)-(0,0.25)$) node [pos=0.3,below] {$\textbf{k}\, \overline{\sigma}$};
                \draw [>=latex, ->, thick, color=black] ($(b)-(0,0.25)$) -- ($(b) + (1,-0.5)$) node [pos=0.7, below] {$\textbf{k'}\, \overline{\sigma}$};
                \draw [thick, densely dashed] ($(b) +(0,0.25)$) -- ($(b)-(0,0.25)$); 
                
                \coordinate (c) at ($(a_cr) + (0,-0.25)$);
                
                \draw [thick, black] ($(c) + (-1,0.5)$) -- ($(c)+(0,0.25)$) node [pos=0.3,above] {$\textbf{p}\, \sigma$};
                \draw [>=latex, ->, thick, color=black] ($(c)+(0,0.25)$) -- ($(c)+(1,-0.5)$) node [shift={(-0.3,0.1)}, below] {$\textbf{k'}\, \sigma$};
                \draw [thick, black] ($(c) - (1,0.5)$) -- ($(c)-(0,0.25)$) node [pos=0.3,below] {$\textbf{k}\, \overline{\sigma}$};
                \draw [>=latex, ->, thick, color=black] ($(c)-(0,0.25)$) -- ($(c) + (1,0.5)$) node [shift={(-0.3,-0.1)}, above] {$\textbf{p'}\, \overline{\sigma}$};
                \draw [thick, densely dashed] ($(c) +(0,0.25)$) -- ($(c)-(0,0.25)$); 
                \node at ($(b) + (0,1.3)$) {spin conserving,}; 
                \node at ($(b) + (0,1.0)$) {opposite spin}; 
                \node at ($(c) + (0,1.3)$) {spin flip}; 
                \node at ($(a_cl) + (0,-0.25) + (-1.1,1.3)$) {b)};
                \node at ($(a_cr) + (0,-0.25) + (-1.1,1.3)$) {c)};
                \node at ($(a_cr) + (-1.1,1.3)$) {};
            \end{tikzpicture}
            \caption{(a) Direct and exchange scattering of two electrons with momentum $\boldsymbol{p}$ and $\boldsymbol{k}$ and spin  $\sigma$, respectively. (b) Spin conserving scattering between opposite spin electrons with spin $\sigma$ and $\overline{\sigma}$. (c) Scattering between opposite spin electrons with spin flip. Adapted from Ref.~\citenum{Penn1985}.}
            \label{fig:scattering_events}
        \end{figure}
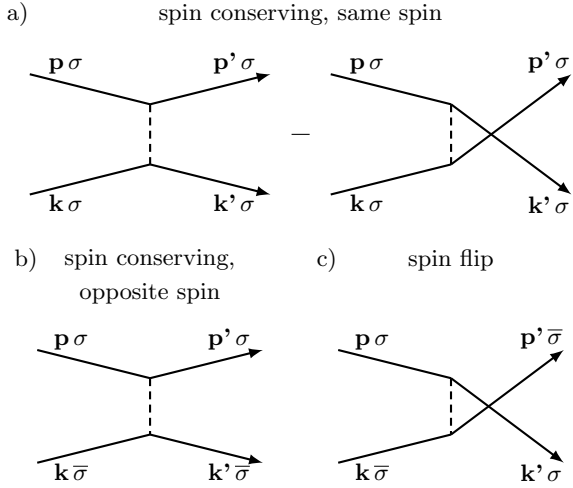     
        In its most general form, the change of the distribution~$f_{\boldsymbol{p}}$ of a given state with momentum $\boldsymbol{p}$ scattering into a state with momentum $\boldsymbol{p'}$ or vice versa
        is given by
        \begin{align}
           \left.\pdv{f_{\boldsymbol{p}}}{t}\right|_{\text{el-el}}
            &= \sum_{\boldsymbol{p'}} \left[\left(1-f_{\boldsymbol{p}}\right)f_{\boldsymbol{p'}} \omega\left(\boldsymbol{p'}, \boldsymbol{p}\right)\right.\nonumber\\
            & \phantom{= \sum_{\boldsymbol{p'}}\left[\right.}- \left.f_{\boldsymbol{p}}\left(1-f_{\boldsymbol{p'}}\right) \omega\left(\boldsymbol{p}, \boldsymbol{p'}\right) \right] \,,
            \label{eq:el_el_scattering_k_general}
        \end{align}
        with the transition probability $\omega(\boldsymbol{p'}, \boldsymbol{p})$ for the scattering into state $\boldsymbol{p}$ and $\omega(\boldsymbol{p}, \boldsymbol{p'})$ for the scattering out of state~$\boldsymbol{p}$, respectively.

        In electron-electron collisions, the momentum of the scattering partner also changes.
        \Cref{fig:scattering_events} shows possible scattering events of an electron of momentum $\boldsymbol{p}$ with an electron of momentum $\boldsymbol{k}$ into states $\boldsymbol{p'}$ and $\boldsymbol{k'}$. 
        Scattering events may occur between electrons with the same spin~$\sigma$ or involve an opposite spin~$\overline{\sigma}$.
        The figure and the notation have been adapted from Ref.~\citenum{Penn1985}.
        Here, we consider a non-magnetic solid; therefore, momenta and energies of the electrons are spin-degenerate and, accordingly, the distributions are independent of spin. 
        
        From Fermi's golden rule, we can determine the transition probability for a scattering event of two electrons, with initial fixed momentum $\boldsymbol{p}$ and any possible $\boldsymbol{k}$ and final states with given momentum $\boldsymbol{p'}$ and any possible $\boldsymbol{k'}$
        by sums over all possible $\boldsymbol{k}$ and $\boldsymbol{k'}$ as
        \begin{align}
            \omega\left(\boldsymbol{p},\boldsymbol{p'}\right) &= \frac{2\pi}{\hbar} \sum_{\boldsymbol{k}\boldsymbol{k'}}f_{\boldsymbol{k}}\left(1-f_{\boldsymbol{k'}}\right)
            \abs{M\left(\boldsymbol{p},\boldsymbol{k},\boldsymbol{p'},\boldsymbol{k'}\right)}^2\nonumber\\
            &\phantom{= \frac{2\pi}{\hbar} \sum_{\boldsymbol{k}\boldsymbol{k'}}\,}\times \delta\left(E_{\boldsymbol{p}} + \varepsilon_{\boldsymbol{k}} - E_{\boldsymbol{p'}} - \varepsilon_{\boldsymbol{k'}}\right)\,,
            %\abs{M\left(\boldsymbol{p'},\boldsymbol{p},\boldsymbol{k'}, \boldsymbol{k}\right)}^2
            \label{eq:transition_probability}
        \end{align}
        where the delta function ensures energy conservation with % is ensured by the delta function, 
        $E_{\boldsymbol{p}}$ and $\varepsilon_{\boldsymbol{k}}$, which denote the energies of the states with momentum $\boldsymbol{p}$ and $\boldsymbol{k}$, respectively.
        The effective matrix element $M\left(\boldsymbol{p},\boldsymbol{k},\boldsymbol{p'},\boldsymbol{k'}\right)$ consists of several terms considering the contributions of the scattering events shown in \cref{fig:scattering_events}. 

        Together, we obtain the change of the distribution $f_{\boldsymbol{p}}$ in \eqref{eq:el_el_scattering_k_general} as
        \begin{align}
            \left.\pdv{f_{\boldsymbol{p}}}{t}\right|_{\text{el-el}} \!\! = \frac{2\pi}{\hbar} \, \frac{1}{2}\sum_{\boldsymbol{p'}\boldsymbol{k}\boldsymbol{k'}} &\delta\left(E_{\boldsymbol{p}} + \epsilon_{\boldsymbol{k}} - E_{\boldsymbol{p'}} - \epsilon_{\boldsymbol{k'}}\right) \nonumber\\
            &\times\left[ \mathcal{F}_{\text{in}} \abs{M\left(\boldsymbol{p'},\boldsymbol{k'},\boldsymbol{p},\boldsymbol{k}\right)}^2 \right.\nonumber \\
            &\phantom{\times \left[\right.}\left.-\, \mathcal{F}_{\text{out}} \abs{M\left(\boldsymbol{p},\boldsymbol{k},\boldsymbol{p'},\boldsymbol{k'}\right)}^2\right]\,,
            \label{eq:el_el_scattering_Mpkpk}
        \end{align}
        with the collision functionals
        \begin{subequations}\label{eq:functionals_momentum}
            \begin{align}
                \mathcal{F}_{\text{in}} &= f_{\boldsymbol{p'}}f_{\boldsymbol{k'}}(1-f_{\boldsymbol{p}})(1-f_{\boldsymbol{k}})\\
                \mathcal{F}_{\text{out}} &= f_{\boldsymbol{p}}f_{\boldsymbol{k}}(1-f_{\boldsymbol{p'}})(1-f_{\boldsymbol{k'}})\, ,
            \end{align}
        \end{subequations}
        representing scattering \textit{into} and \textit{out of} the state with momentum $\boldsymbol{p}$, respectively.
        The factor $1/2$ enters to avoid double counting of the momenta $\boldsymbol{p'}$ and $\boldsymbol{k'}$ \cite{Penn1985}.
        
    \subsection{Scattering terms}\label{subsec:scattering terms}
        \Cref{fig:scattering_events} sketches all considered scattering events out of state $\boldsymbol{p}$.
        These are summarized in one effective matrix element in \cref{eq:el_el_scattering_Mpkpk}. 
       It consists of the contributions
        \begin{align}
            \abs{M\left(\boldsymbol{p},\boldsymbol{k},\boldsymbol{p'},\boldsymbol{k'}\right)}^2 &= \abs{M_{\boldsymbol{p'}\sigma,\boldsymbol{k'}\sigma}^{\boldsymbol{p}\sigma,\boldsymbol{k}\sigma} - M_{\boldsymbol{k'}\sigma,\boldsymbol{p'}\sigma}^{\boldsymbol{p}\sigma,\boldsymbol{k}\sigma}}^2\nonumber\\
            &\phantom{=\,\,} +  \abs{M_{\boldsymbol{p'}\sigma,\boldsymbol{k'}\overline{\sigma}}^{\boldsymbol{p}\sigma,\boldsymbol{k}\overline{\sigma}}}^2 +  \abs{M_{\boldsymbol{k'}\sigma,\boldsymbol{p'}\overline{\sigma}}^{\boldsymbol{p}\sigma,\boldsymbol{k}\overline{\sigma}}}^2 . 
            \label{eq:sum_matrix_elements}
        \end{align}
        Here, the superscript indices refer to the initial momenta before the collision, while the subscript indices indicate the momenta after the collision.
        \Cref{eq:el_el_scattering_Mpkpk} also accounts for the analogous time-reversal scattering events into state $\boldsymbol{p}$ with initial states $\boldsymbol{p'}$ and $\boldsymbol{k'}$.
        
        Each of the individual matrix elements  in \cref{eq:sum_matrix_elements} is determined through the same scattering potential $V_{\text{e}\text{e}}$ as exemplarily shown for
        \begin{align}        
            M_{\boldsymbol{p'}\sigma,\boldsymbol{k'}\overline{\sigma}}^{\boldsymbol{p}\sigma,\boldsymbol{k}\overline{\sigma}} &= \bra{\boldsymbol{p'}\sigma,\boldsymbol{k'}\overline{\sigma}} V_{\text{e}\text{e}} \ket{\boldsymbol{p}\sigma,\boldsymbol{k}\overline{\sigma}}\label{eq:general_matrix_element} \,.
        \end{align}
       
        The first term in \eqref{eq:sum_matrix_elements} is a difference of matrix elements, since the processes in \cref{fig:scattering_events} a) are not distinguishable and the total scattering amplitude for Fermions has to be antisymmetric under exchange of two identical particles.
        In our work, however, we neglect the interference term, hence
        \begin{align}
            \abs{M_{\boldsymbol{p'}\sigma,\boldsymbol{k'}\sigma}^{\boldsymbol{p}\sigma,\boldsymbol{k}\sigma} - M_{\boldsymbol{k'}\sigma,\boldsymbol{p'}\sigma}^{\boldsymbol{p}\sigma,\boldsymbol{k}\sigma}}^2 &= \abs{M_{\boldsymbol{p'}\sigma,\boldsymbol{k'}\sigma}^{\boldsymbol{p}\sigma,\boldsymbol{k}\sigma}}^2 + \abs{M_{\boldsymbol{k'}\sigma,\boldsymbol{p'}\sigma}^{\boldsymbol{p}\sigma,\boldsymbol{k}\sigma}}^2\nonumber\\
            &\phantom{=\,\,} -  \left(M_{\boldsymbol{p'}\sigma,\boldsymbol{k'}\sigma}^{\boldsymbol{p}\sigma,\boldsymbol{k}\sigma}{M_{\boldsymbol{k'}\sigma,\boldsymbol{p'}\sigma}^{\boldsymbol{p}\sigma,\boldsymbol{k}\sigma}}^{*} + c.c.\right) \nonumber \\
            &\approx \abs{M_{\boldsymbol{p'}\sigma,\boldsymbol{k'}\sigma}^{\boldsymbol{p}\sigma,\boldsymbol{k}\sigma}}^2 + \abs{M_{\boldsymbol{k'}\sigma,\boldsymbol{p'}\sigma}^{\boldsymbol{p}\sigma,\boldsymbol{k}\sigma}}^2\,,\label{eq:interference_term_neglected}
        \end{align}
        resulting in the approximation
        \begin{align}
            \abs{M\left(\boldsymbol{p},\boldsymbol{k},\boldsymbol{p'},\boldsymbol{k'}\right)}^2 & \approx 
            \abs{M_{\boldsymbol{p'}\sigma,\boldsymbol{k'}\sigma}^{\boldsymbol{p}\sigma,\boldsymbol{k}\sigma}}^2 + \abs{M_{\boldsymbol{k'}\sigma,\boldsymbol{p'}\sigma}^{\boldsymbol{p}\sigma,\boldsymbol{k}\sigma}}^2\nonumber\\
            &\phantom{=\,} +  \abs{M_{\boldsymbol{p'}\sigma,\boldsymbol{k'}\overline{\sigma}}^{\boldsymbol{p}\sigma,\boldsymbol{k}\overline{\sigma}}}^2 +  \abs{M_{\boldsymbol{k'}\sigma,\boldsymbol{p'}\overline{\sigma}}^{\boldsymbol{p}\sigma,\boldsymbol{k}\overline{\sigma}}}^2,
            \label{eq:sum_matrix_elements_approx}
        \end{align}
        which is equivalent to assuming that all four processes sketched in \cref{fig:scattering_events} are distinguishable. 

        We assume that all of these matrix elements are the same, and we denote them as $M_{\text{ee}}$. 
        They are determined from a screened Coulomb potential, also known as Yukawa potential, 
        \begin{equation}
            V_{\text{ee}} = \frac{e^2}{4\pi\varepsilon_0 r}\e{-\kappa r}\,,
            \label{eq:coulomb_potential}
        \end{equation}
        with the screening parameter $\kappa$. 
        \Cref{eq:general_matrix_element} then results in~\cite{Ziman_electrons_phonons,Gasparov1993}
        \begin{equation}
            \abs{M_{\text{ee}}(\Delta k)}^2 = \left( \frac{e^2}{\varepsilon_0\Omega_0} \frac{1}{\Delta k^2 + \kappa^2} \right)^2 \,,
            \label{eq:coulomb_matrix_element}
        \end{equation}
        where $\Omega_0$ denotes the volume of the unit cell and 
        \begin{equation}
                \Delta k^2 = (\boldsymbol{k} - \boldsymbol{k'})^2 = (\boldsymbol{p} - \boldsymbol{p'})^2 \,,
        \end{equation}
        describes the exchanged momentum.

    \subsection{Random-$\boldsymbol{k}$ approximation}
        Within the random-$\boldsymbol{k}$ approximation, any specific angular dependence can be disregarded, as it is assumed to be rapidly smeared out by elastic scattering~\cite{Penn1985}.
        Thus, we obtain spherically-averaged quantities by transforming the sums over momenta in \eqref{eq:el_el_scattering_Mpkpk} into energy integrals with~\cite{Knorren2000}
        \begin{equation}
            \sum_{\boldsymbol{k}} f_{\boldsymbol{k}} \rightarrow \niceint{\varepsilon}D_{\varepsilon}f_{\varepsilon}\,,
            \label{eq:random_k_transformation}
        \end{equation}
        where the electronic density of states $D_{\varepsilon}$ is introduced.
        The energy-dependent distribution $f_{\varepsilon}$ is the spherical average of the momentum-dependent distribution $f_{\boldsymbol{k}}$.
        This transformation is strictly valid only for electrons, for which the relation
        \begin{equation}
            D_{\varepsilon} = \frac{\Omega_0}{2\pi^2} k^2 \dv{k}{\varepsilon}
            \label{eq:oneband_condition}
        \end{equation}
        holds, as, for example,  for free electrons with the dispersion relation $\varepsilon \propto k^2$.
        
        The same transformation,  \eqref{eq:random_k_transformation}, is applied to all sums in \eqref{eq:el_el_scattering_Mpkpk}, performing the spherical averages of the participating momenta and replacing them with the corresponding energy values, that is \mbox{$\boldsymbol{p} \rightarrow E$}, \mbox{$\boldsymbol{p'} \rightarrow E'$}, \mbox{$\boldsymbol{k} \rightarrow \varepsilon$} and \mbox{$\boldsymbol{k'} \rightarrow \varepsilon'$}.
        The energy-conserving delta function allows us to evaluate the integral over $\varepsilon'$ by assigning $\varepsilon'=\varepsilon + \Delta E$ with $\Delta E = E-E'$.
        Note that the random-$\boldsymbol{k}$ approximation includes the assumption that momentum conservation is implicitly fulfilled. 

        To complete the transformation, we also need to determine energy-dependent matrix elements, $M_{\text{ee}}(\Delta E)$, by spherical averages of the momentum-dependent matrix elements.
        In subsection \ref{subsec:matrixelement}, we will go one step further and average the matrix elements over all possible exchanged momenta, ending up with $M_{\text{ee}}$ being independent of momentum or energy. 
        
        The random-$\boldsymbol{k}$ approximation, with the transformations according to \eqref{eq:random_k_transformation}, the implicit momentum conservation and spherical averages of all momentum-dependent quantities and functions 
        leads to the final expression for the full Boltzmann collision integral
        \begin{align}
            \left.\pdv{f_{E}}{t}\right|_{\text{el-el}}&= \frac{2\pi}{\hbar} \, g \niceint{E'}\niceint{\varepsilon} D_{E'}D_{\varepsilon}D_{\varepsilon + \Delta E}\mathcal{F}_E\abs{M_{\text{ee}}}^2 \,, 
            \label{eq:energy-dependent_collision_integral}
        \end{align}
        with the energy-dependent collision functional
        \begin{align}
            \mathcal{F}_E &= f_{E'}f_{\varepsilon+\Delta E}(1-f_{E})(1-f_{\varepsilon})\nonumber \\
            &\phantom{=\,\,}- f_{E}f_{\varepsilon}(1-f_{E'})(1-f_{\varepsilon+\Delta E}) \,,
        \end{align}
        and the spin-degeneracy factor $g\!=\!2$, here entering through the two last terms in \eqref{eq:sum_matrix_elements_approx} describing the processes depicted in \cref{fig:scattering_events} b) and c).
        
        Energy-dependent collision integrals of this form 
        have been applied  
        to study ultrafast dynamics in noble metals, ferromagnets, and semiconductors~\cite{Knorren2000,Mueller2013PRB,Zarate1999,Penn1985,Snoke1992,Seibel2023,Seibel2025CommPhys}.
    
    \subsection{Matrix element}
    \label{subsec:matrixelement}

        In the framework of the random-$\boldsymbol{k}$ approximation, we lose information about the exchanged momentum in a collision event, consequently a spherically averaged matrix element needs to be applied in the collision integral (\ref{eq:energy-dependent_collision_integral}). 
        Here, we additionally 
        consider the energy-dependence of the matrix element to be negligible, meaning that it is the same for all scattering events.
        Therefore, we average the matrix element from \eqref{eq:coulomb_matrix_element} over all $k$ and $k'$ in the unit sphere
        \begin{align}
            \expval{M_{\text{ee}}^2} &= \frac{\Omega_0^2}{(2 \pi)^6} \niceint{^3 k}{\niceint{^3 k'}{ \abs{M_{\text{ee}}(\Delta k)}^2} }\nonumber\\
            &= \frac{e^4}{16 \pi^4 \varepsilon_0^2} \left[ \frac{k_r^3}{3\kappa}\left( \frac{\pi}{2} + \arctan{\frac{4k_r^2 - \kappa^2}{4k_r\kappa}}\right)\right.\nonumber\\
            &\phantom{=\,}+\left. \left(\frac{k_r^2}{2} + \frac{\kappa^2}{12} \right) \ln\left(\frac{\kappa^2}{4k_r^2 + \kappa^2}\right) + \frac{k_r^2}{3} \right] ,
            \label{eq:coulomb_avg}
        \end{align}
        where $k_r = \sqrt[3]{\frac{6\pi^2}{\Omega_0}}$ is the radius of the unit sphere in momentum space.
        
        The averaged Coulomb matrix element is independent of the electronic momentum or energy, and depends only on the screening.
        The screening parameter $\kappa$ is calculated at each timestep
        in dependence on the electron distribution~\cite{DelFatti2000,Mueller2013PRB,Binder1997}
        \begin{equation}
            \kappa^2(t) = \frac{e^2}{\varepsilon_0} \niceint{E} D_E \dv{f_E(t)}{E},
        \end{equation}
        which enables the consideration of thermal broadening and non-equilibrium effects.
        
    \subsection{Relaxation time approach\label{subsec:relaxation_time_approach}}
        The relaxation time approach (RTA) is often used as a computationally simple alternative to the full collision integral (FCI) for capturing the electron thermalization.
        In the RTA, the change of the distribution is given by 
        \begin{align}
            \left.\pdv{f_{E}(t)}{t}\right|_{\text{el-el}} = - \frac{f_E(t)-f_E^{\text{Fermi}}}{\tau}\,,
            \label{eq:relaxation_time_approach_general}
        \end{align}
        with a relaxation time $\tau$.
        Here, $f_E(t)$ denotes the time-dependent non-equilibrium electron distribution, while $f_E^{\text{Fermi}}$ is the unique Fermi distribution with energy and particle density of the thermalized equilibrium state which defines also the corresponding quasi-temperature of the electron system, $T_\text{e}$.

        Formally, only a constant relaxation time  ensures the conservation of energy and particle density during thermalization \cite{Uehlein2025}.
        However, for weak excitations, the relaxation time is often identified with the lifetime of an excited electron, which is known to be strongly dependent on the energy state $E$ of the excited electron \cite{Bauer2015}.
        The Fermi liquid theory allows the calculation of this single-electron lifetime, resulting in \cite{Mueller2013PRB,Pines,Kaveh1984}
        \begin{align}
            \tau_\text{E} = \tau_0 \frac{E_\text{F}^2}{\left(E-E_\text{F}\right)^2 + \left(\pi k_\text{B}T_\text{e}\right)^2} \,,
            \label{eq:relaxation_time_energy_dependent}
        \end{align}
        with the Fermi energy $E_\text{F}$ and a material-dependent \mbox{pre-factor} $\tau_0$, theoretically determined by
        \begin{align}
            \tau_{0,\,\text{calc}} = \frac{128}{\sqrt{3}\pi^2\omega_\text{p}}\,,
            \label{eq:tau0}
        \end{align}
        where $\omega_\text{p}$ is the plasma frequency \cite{Carpene2006,Pines}.

        In our analysis, we further determine a modified value, $\tau_{0,\,\text{fit}}$, from a comparison with the outcome of the kinetic simulation applying the full Boltzmann collision integral (FCI) and study the resulting thermalization dynamics for the RTA (\ref{eq:relaxation_time_approach_general}) with $\tau \!=\!\tau_E$ from \eqref{eq:relaxation_time_energy_dependent} with both expressions for $\tau_0$. 
        Moreover, we examine the thermalization resulting from the RTA with a constant relaxation time, $\tau \!=\! \tau_{\text{const}}$, which we also determine from a fit to the kinetic approach including the FCI. 
    
    \section{Results\label{sec:results}}
    In this work, we compare the accuracy and benefits of kinetic Boltzmann simulations using the full collision
    integral (FCI) and the relaxation time approach (RTA) to describe the thermalization of electrons. 
    We use the \mbox{random-$\boldsymbol{k}$} approximation for the FCI calculations with \eqref{eq:energy-dependent_collision_integral} and we study different models for the relaxation time $\tau$ in \eqref{eq:relaxation_time_approach_general} for the RTA. 
    The following results are calculated for a free electron gas (FEG) with $D(E) \propto \sqrt{E}$ for which the random-$\boldsymbol{k}$ approximation holds. 
    
    \subsection{Initial distribution}\label{subsec:initialdistribution}
        We focus on the thermalization of the electrons. Therefore, we initialize all calculations with the same non-equilibrium distribution.
        It is calculated using a full Boltzmann collision integral describing the excitation with optical photons, as given in Refs.~\citenum{Rethfeld2002,Mueller2013PRB,Seibel2023}. We consider a laser wavelength of \SI{400}{\nm}
        corresponding to a photon energy of $3.1\,$eV. 
        We choose an absorbed energy which leads to an internal energy of the electron system corresponding to a \mbox{(quasi-)}temperature of about 1800\,K.
        
\begin{figure}[b]
            \centering
            \includegraphics[scale=1]{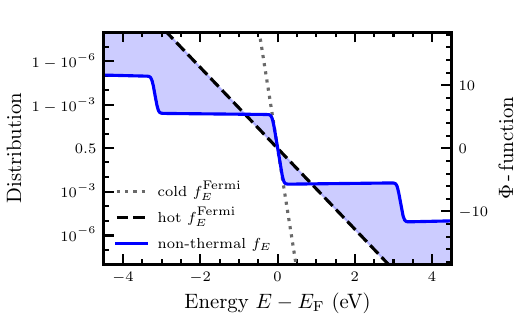}
            \caption{Step-shaped, non-thermal distribution of an excited, degenerate free-electron gas (blue solid line) together with its corresponding hot Fermi distribution (dashed line) as well as a cold Fermi distribution at 300\,K (dotted line). The distribution is plotted in a symmetric logarithmic representation know as the $\Phi$-function introduced in Ref.~\citenum{Rethfeld2002} (right y-axis) or the mathematically equivalent logit-function (left y-axis). In this plot, Fermi functions are represented as a linear function with a slope inversely proportional to temperature, and deviations from a straight line visualize the non-equilibrium.}
            \label{fig:illustration_deviation}
        \end{figure}

        \Cref{fig:illustration_deviation} shows the laser-excited electronic energy distribution before thermalization.
        The excitation results in a step-shaped, non-thermal distribution.
        The steps are observed above and below the Fermi edge with the width equal to the photon energy. Thus, the curvature of the Fermi edge is reproduced at $\pm 3.1\,$eV.
        The distribution is plotted in a symmetric logarithmic representation known as $\Phi = \ln{(f^{-1} -1)}$ \cite{Rethfeld2002}, or the mathematically equivalent logit-function. 
        In both cases, a Fermi distribution corresponds to a linear dependence on energy with a slope antiproportional to the temperature. 
        The so-called corresponding hot Fermi distribution with the same energy density as the non-thermal, excited distribution is therefore less steep than the cold Fermi distribution at room temperature before excitation. 
        In this representation, the deviation of the non-thermal, excited distribution $f_E$ from a straight line directly visualizes the deviation from a Fermi distribution.
        
    \subsection{Relaxation times}
        First, we extract a constant relaxation time $\tau_\text{const}$ for the RTA from our Boltzmann calculation using FCI, which allows us to better compare both methods. Therefore, we calculate the temporal evolution of the mean absolute deviation (MAD)
        \begin{align}
        	\Delta_{\text{MAD}}(t)=\niceint{E}\,\abs{f_E(t)-f_E^\text{Fermi}}
            \label{eq:MAD}
        \end{align}
        of the distribution $f_E(t)$ from equilibrium. The deviation between the two distributions in \eqref{eq:MAD} is illustrated in \cref{fig:illustration_deviation} by the colored area between the non-thermal distribution $f_E(t\!=\!0\,\mathrm{fs})$ and the hot Fermi distribution~$f_E^{\text{Fermi}}$.
        
\begin{figure}[t]
        	\centering
        	\includegraphics[width=1\linewidth]{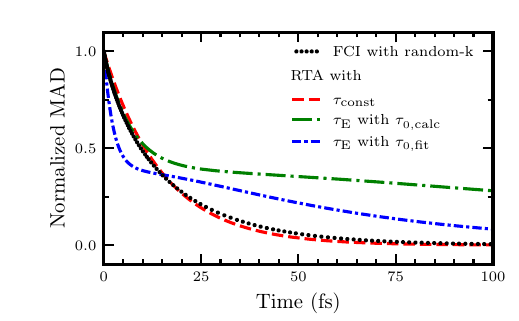}
        	\caption{Normalized MAD of the non-thermal distribution from equilibrium for the FCI with random-$\boldsymbol{k}$ (black, dotted line) compared to three types of the RTA: constant $\tau_{\text{const}}$ (red, dashed line), energy-dependent $\tau_\text{E}$ with $\tau_{0,\text{calc}}$ (green, long dash-dotted line), and energy-dependent $\tau_\text{E}$ with $\tau_{0,\text{fit}}$ (blue, short dash-dotted line). An exponential fit provides the same fitted thermalization time of $15.2\,$fs for all curves except for the green one which relaxes with a characteristic time of $39.8\,$fs.} 
        	\label{fig:MAD_all_models}
        \end{figure}

        \Cref{fig:MAD_all_models} shows the decay of the normalized MAD for the FCI calculation over time due to the thermalization of the electron system.
        This FCI curve, fitted with
        \begin{align}
        	\Delta_{\text{MAD},\text{Fit}}(t)= \text{exp}\left(-\frac{t}{\tau_\text{const}}\right)\,,
        \end{align}
        provides $\tau_\text{const}=15.2\,$fs for the considered system.
        The curve resulting from the RTA with this constant time looks qualitatively and quantitatively similar to the FCI curve.

        Secondly, we analyze the energy-dependent relaxation time based on the Fermi liquid theory in \eqref{eq:relaxation_time_energy_dependent} and use the plasma frequency $\hbar\omega_\text{p}= 17.34\,$eV calculated based on the effective mass $m=1.65\, m_\text{e}$ and the Fermi energy $E_{\text{F}}=11.18\,$eV \cite{phd_Mueller}. 
        With this plasma frequency, the pre-factor calculates to $\tau_{0,\,\text{calc}}=\SI{0.284}{fs}$.
        The results for this RTA are also shown in \cref{fig:MAD_all_models} (green, dash-dotted line).
        In contrast to the RTA with constant relaxation time (red) and the calculation with FCI (black), a slower thermalization can be observed.
        A fit to these data provides a characteristic time of $39.8\,$fs for the decay of the MAD, which exceeds the characteristic time of $15.2\,$fs for the other two models mentioned before.
        
        Due to this discrepancy we introduce a third approach to determine the relaxation time also using \eqref{eq:relaxation_time_energy_dependent} but with an adapted pre-factor $\tau_{0,\text{fit}}=\SI{0.1}{fs}$.
        It is used as a fit parameter to force the MAD of this approach to yield the same fitted thermalization time as for the calculation with FCI.
        The resulting normalized MAD is 
        shown in \cref{fig:MAD_all_models} with the blue dash-dotted lines.
        Despite the now identical fitted thermalization times for these approaches, deviations in the temporal evolution of the MAD persist.
        Within the first $15\,$fs, the RTA with energy-dependent $\tau_E$ with $\tau_{0,\text{fit}}$ thermalizes faster as compared to the FCI, while on longer timescales it thermalizes slower. 
        To further investigate the details of the thermalization, the energy-resolved distributions are analyzed.
    
    \subsection{Energy-resolved distribution}
        \begin{figure*}[ht]
        	\centering
            	\includegraphics[scale=1]{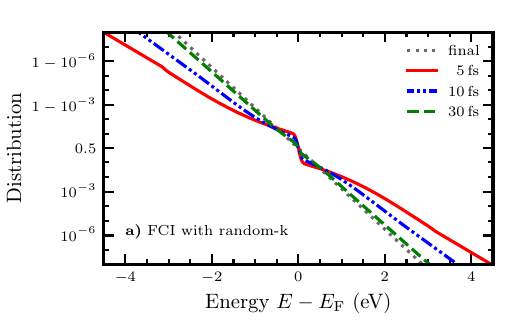}
        		\includegraphics[scale=1]{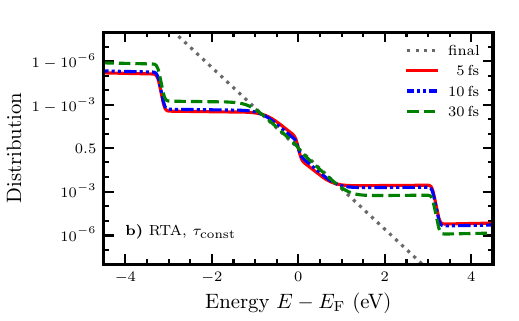}
        	\vspace{0.8em}
        	    \includegraphics[scale=1]{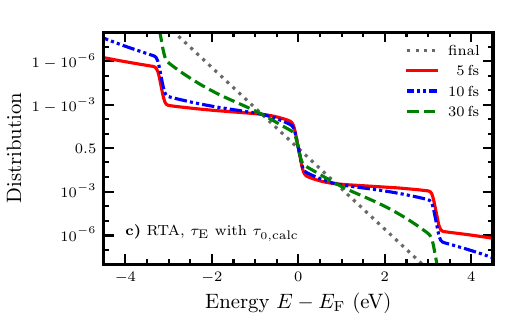}
        	    \includegraphics[scale=1]{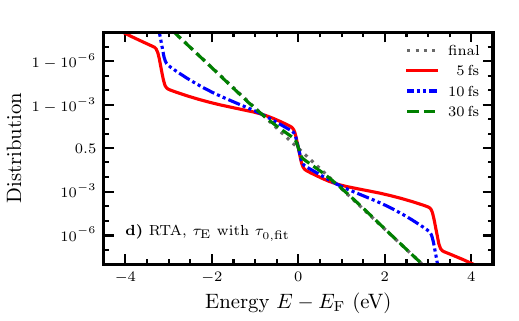}
        	\caption{Thermalization of the distribution after laser excitation. a) Calculated with FCI with random-$\boldsymbol{k}$, b) to d) calculated with RTA with different relaxation times: constant $\tau_{\text{const}}$ (b), energy-dependent $\tau_\text{E}$ with $\tau_{0,\text{calc}}$ (c), and energy-dependent $\tau_\text{E}$ with $\tau_{0,\text{fit}}$ (d).}
            \label{fig:distribution_at_different_times}
        \end{figure*}
        We analyze the effects of the description of the electron-electron interaction on the thermalization of the energy distribution of excited electrons.
        \Cref{fig:distribution_at_different_times} shows the energy-resolved distribution at different times after laser excitation in the symmetric logarithmic representation
        for four different assumptions. 
        Results for a calculation with FCI in the \mbox{random-$\boldsymbol{k}$} approximation according to \eqref{eq:energy-dependent_collision_integral} are shown in \cref{fig:distribution_at_different_times} a), and the results of the various approaches applied for the RTA according to \eqref{eq:relaxation_time_approach_general} are shown in \cref{fig:distribution_at_different_times} b)-d).   
        Note again, that the initial non-thermal distribution for all calculations is the same, depicted in \cref{fig:illustration_deviation}.

        \Cref{fig:distribution_at_different_times} a) shows a rapid softening of the initial step structure. 
        The signature of the cold Fermi edge is still visible after $5\,$fs  or $10\,$fs, while its reproductions at $3.1\,$eV above and below $E_\text{F}$ are no longer present and the plateaus of the photon absorption have disappeared. 
        The overall slope of the distribution resembles a hotter energy distribution than the final Fermi distribution suggests. 
        Therefore, possible measurements of highly-excited electrons may overestimate the temperature of the system~\cite{Medvedev2011,Zastrau2008,Vinko2010etal}.
        Within about $30\,$fs, the thermalizing distribution closely resembles the final equilibrium state.
    
        The calculation with the RTA applying a constant relaxation time, \cref{fig:distribution_at_different_times} b), shows a rather different thermalization dynamics.
        Here, the characteristic step-structure of the initial excitation is preserved over all evaluated times. 
        In the vicinity of the Fermi edge, the thermalization appears to be faster than further away for high-energy electrons (or low-energy holes, respectively). 
        While the optical impression is partially a consequence of the logarithmic representation, the distribution clearly overestimates the hot tail of the excited distribution. 
        Evaluations referring to these electrons will be strongly different from the calculation with the full Boltzmann collision integral (FCI). 

        Energy-dependent relaxation times can to some extend heal this deviation. 
        \Cref{fig:distribution_at_different_times} c) and d) show the results of the calculations with the energy-dependent relaxation time based on Fermi liquid theory, according to \eqref{eq:relaxation_time_energy_dependent}.
        In the high-energy region, the thermalization towards the final hot Fermi distribution is much faster than for the calculation with a constant relaxation time,
        while around the Fermi edge, the signature of the cold edge is preserved for longer time. 
        Both aspects rather resemble the dynamics obtained with the FCI, though around Fermi edge and at the step-edges at $\pm 3.1\,$eV, the non-equilibrium structure remains for longer times than in the calculation with the FCI. 

        In total, \cref{fig:distribution_at_different_times} reveals that the RTA with an energy-dependent relaxation time and a fitted prefactor $\tau_{0,\text{fit}}$ leads to a thermalization dynamics where the overall distribution with its characteristic features for electrons close to the Fermi edge as well as the amount of highly energetic electrons and holes show the best resemblance to the calculation applying the FCI, compare panels (d) and (a). 
        It particularly also reproduces the nearly-complete thermalization at a time of $t=30\,$fs after the initial excitation. 
        Thus, the evaluation of an integrated quantity like the mean average deviation (MAD), see \cref{fig:MAD_all_models}, is not sufficient to judge about the accuracy of different models, particularly when the details of the electronic energy distribution matter.
                
    \subsection{Occupation time}    
\begin{figure}[t]
        	\centering
            \includegraphics[width=\linewidth]{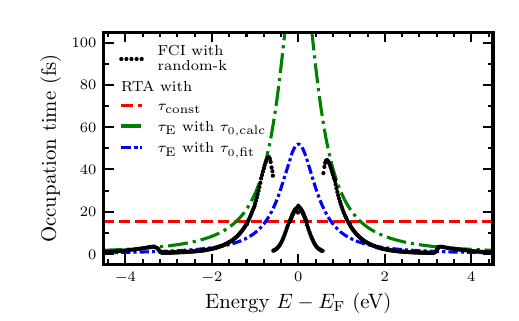}
        	\caption{Energy-dependent occupation times for the FCI with random-$\boldsymbol{k}$ (black) and the RTA, compared for different relaxation times: constant $\tau_{\text{const}}$ (red), energy-dependent $\tau_\text{E}$ with $\tau_{0,\text{calc}}$ (green), and energy-dependent $\tau_\text{E}$ with $\tau_{0,\text{fit}}$ (blue).}
        	\label{fig:occupation_times}
        \end{figure}

        We examine the energy-resolved occupation time of electron states, which includes  scattering out of states as well as their reoccupation.
        To determine this time for the Boltzmann calculation with FCI, the temporal decay of the MAD is fitted, in the same way as for the thermalization time in \cref{fig:MAD_all_models}.
        Here, however, the MAD is fitted with spectral resolution for individual energies, rather than integrating over all energies.
        \Cref{fig:occupation_times} shows the resulting occupation time in dependence on energy for the applied descriptions of electron-electron interaction. 
        The occupation time is symmetric to the Fermi energy, regardless of whether the FCI or the RTA is used, which implies the same rate of thermalization for electrons and holes. Note that this may be different for materials with more complex densities of states \cite{Seibel2025CommPhys}.
        For the models with RTA, the occupation time equals the relaxation time by construction, thus 
        we show the respective relaxation time $\tau$ applied in  \eqref{eq:relaxation_time_approach_general}. 

        The occupation time calculated using Boltzmann with FCI varies significantly depending on the energy.
        It exhibits a local maximum at the Fermi energy, which then decays symmetrically for increasing and decreasing energies. 
        In this region, it qualitatively resembles the behavior of the occupation times resulting from Fermi liquid theory, however, the FCI time is much smaller. 
        Towards $1\,$eV above and below $E_\text{F}$, we find a discontinuity in the occupation times followed by the global maxima. For energies further away from the Fermi edge, the occupation time decreases, apart from two small kinks at $3.1\,$eV distance from $E_\text{F}$.
        A faster thermalization for higher energies is well known for optically excited metals and has also been experimentally observed~\cite{Faure2013, Ruffing2013,Bauer2015}.
        The peaks centered around Fermi energy stem from the intersections of the non-thermal distribution with the corresponding hot Fermi distribution, compare \cref{fig:illustration_deviation},
        where the occupation therefore already equals the thermalized value.
        Note that the energy of these states depend on the excitation conditions like wavelength and excitation strength~\cite{Seibel2023}.        
        Here, the scattering processes into and out of these states balance each other out, thus the occupation remains almost constant ~\cite{Seibel2023}.

        The relaxation times applied in the RTA are different from the occupation time resulting from the ensemble calculation of the FCI in \cref{fig:occupation_times}. 
        The constant relaxation time $\tau_\text{const}$ is smaller in an 
        energetic region around the Fermi edge, while it is considerably longer for higher energies. 
        This explains the differences observed in the dynamics of the nonthermal distribution, see \cref{fig:distribution_at_different_times}.
    
        The applied energy-dependent relaxation time $\tau_E$ from \eqref{eq:relaxation_time_energy_dependent} contains a decay for energies away from Fermi edge as $\propto (E-E_F)^{-2}$.
        As pointed out in \cref{subsec:relaxation_time_approach}, Fermi liquid theory describes the lifetime of individual excited electrons. 
        Therefore, it does not capture any state-filling effects which are responsible for the complex occupation time obtained with the FCI. This is a fundamental shortcoming of the Fermi liquid theory when being applied to the description of ensemble dynamics. 
        Nevertheless, the energy-dependent scattering time in \cref{eq:relaxation_time_energy_dependent} with the adjusted prefactor $\tau_{0,\text{fit}}$ represents a reasonable average around Fermi edge and a good quantitative agreement in the high-energy region.
    
\section{summary and conclusion\label{sec:conclusion}}

In this work, we studied the thermalization of a laser-excited non-equilibrium electron distribution.
We presented a derivation for the full electron-electron collision integral within the random-$\boldsymbol{k}$ approximation, which served as the basis of our analysis.
This framework enables a detailed description of the temporal evolution of the electronic distribution in optically excited Fermi systems.

For comparison, we also examined the dynamics resulting from a relaxation-time approach with different expressions for the relaxation time.
This approach reduces the complexity of microscopic interactions to an effective parameter and is, therefore, commonly used.
Our analysis shows that a constant relaxation time adequately captures the rapid thermalization of electrons near the Fermi energy but fails to describe the dynamics at higher or lower energies.
In contrast, an energy-dependent relaxation time provides significantly better agreement with the full collision-integral results, particularly in the high-energy regime.
This could be especially important in calculations where energy-resolution is important, for transport effects or energy-resolved interface scattering.

We have extracted the energy-resolved occupation times of the full collision integral calculation and compared them to the relaxation times. 
We find that for a free-electron like density of states the dynamics of electrons above and holes below Fermi energy are nearly equivalent. 
We have found that in a few-eV distance from Fermi energy, there is in good agreement with the energy-dependent relaxation time resulting from Fermi liquid theory.
However, the details of the ensemble characteristics, such as a significant increase of occupation time for energies closer to the Fermi energy, due to a balance between scattering and reoccupation can only be captured with a full collision integral. 

    \bibliographystyle{IEEEtran}            
	\bibliography{bibfile/all.bib}
\end{document}